\documentclass[11pt,twoside]{book}
\usepackage{konkolyproc2}
\usepackage{longtable}
\usepackage{amsmath,amssymb}
\usepackage{graphicx}
\usepackage{lscape}
\usepackage{index}
\usepackage{natbib}
\usepackage{bigdelim}
\usepackage{multirow}
\makeindex

\begin{document}

%%%%%%%%%%%%%%%%%%%%%%%%%%%%%%%%%%%%%%%%%%%%%%%%%%%%%%%%%%%%%%%%%%%%%%%%%%
\pagestyle{myheadings}
\setcounter{equation}{0}\setcounter{figure}{0}\setcounter{footnote}{0}\setcounter{section}{0}\setcounter{table}{0}\setcounter{page}{1}
\markboth{Szabados, Szab\'o \& Kinemuchi}{RRL2015 Conf. Papers}
\title{Review of candidates of binary systems with an RR Lyrae component}
\author{Marek Skarka$^{1,2}$, Ji\v{r}\'{i} Li\v{s}ka$^{2}$, Miloslav Zejda$^{2}$ \& Zden\v{e}k Mikul\'{a}\v{s}ek$^{2}$}
\affil{$^{1}$Konkoly Observatory, Research Centre for Astronomy and
Earth Sciences, Hungarian Academy of Sciences, H-1121 Budapest, Konkoly Thege Mikl\'{o}s \'{u}t 15-17, Hungary, marek.skarka@csfk.mta.hu}
\affil{$^{2}$Department of Theoretical Physics and Astrophysics, Masaryk University, Brno, Czech Republic}%, \texttt{maska@physics.muni.cz}}
%%%%%%%%%%%%%%%%%%%%%%%%%%%%%%%%%%%%%%%%%%%%%%%%%%%%%%%%%%%%%%%%%%%%%%%%%%%

\begin{abstract}
We present an overview and current status of research on RR~Lyrae stars in binary systems. In present days the number of binary candidates has steeply increased and suggested that multiple stellar systems with an RR Lyrae component is much higher than previously thought. We discuss the probability of their detection using various observing methods, compare recent results regarding selection effects, period distribution, proposed orbital parameters and the~Blazhko effect.% and introduce another candidates for RR Lyrae stars bound in binary systems.
\end{abstract}
\vspace*{-0.8cm}
\section{Introduction}\label{Sec:Introduction}

	It is generally assumed that majority of stars resides binary or even multiple systems. However, the situation is not so simple because the estimates of their incidence differ for different stellar populations and stellar types. For example, \citet{sana2011} give about 40\,\% of O and B type stars, \citet{duquennoy1991} give 60--80\% of F and G stars, \citet{lada2006} proposes that 30\,\% of stars of all stellar types are bound in binaries.
	
	Nevertheless, we know that many pulsating stars really orbit around common center of mass with some kind of companion. There are more than 150 cepheids and more than 100 $\delta$\,Sct type stars known in binaries \citep{szabados2003,liakos2012}. The list assembled from available literature is provided by \citet{zhou2014}.

	What is the situation with RR Lyrae stars? There are only 61 candidates known so far\footnote{And several tens of candidates in globular clusters \citep[e.g.][]{jurcsik2012} and in the Galactic Bulge \citep[][this proceedings]{hajdu2015b}.} \citep[][this proceedings]{liska2015c}, and only one system, in which a pulsating component is not a classical RR~Lyrae, has been confirmed \citep{pietrzynski2012}. Considering the fraction of known to all binary RR~Lyrae candidates this is less than 0.1\,\%. The reasons for this unpleasant situation emerge mainly from stellar evolution producing difficulties in detection of binarity. 

\section{Expected characteristics of binary candidates}\label{Sec:Characteristics}
\vspace*{-0.2cm}
	Binary system with RR Lyrae component should be well detached because otherwise mass transfer causing different evolutionary scenario could take place. Such wide binary with orbital period longer than a few hundreds of days will be hardly detectable because of very low probability of eclipses, low amplitude of radial velocity (hereafter RV) variations vanishing in RV changes caused by pulsations. Difficulties in detection are also caused by the necessity of long-term monitoring which is not always available, or possible to do.
	
	If the initial mass of the companion was higher than of the RR Lyrae component it would evolve much faster and should presently be in a form of degenerated remnant -- either white dwarf, neutron star, or a black hole. Such binaries would definitely not be detectable as eclipsing and only spectral lines of RR Lyrae component would be visible (SB1 type). If the companion evolved faster, then the RR Lyrae component could possibly be contaminated with heavier elements originated form the ejected envelope during the last stadia of more massive companion \citep[e.g.][]{kennedy2014}.
	
	Concerning a low-mass companion, it could be in all evolutionary stadia. Radius and luminosity of the companion will influence to what extend it will manifest itself in observational manner:
	\begin{itemize}
		\item {\it Main sequence star} -- amplitude of eclipses would be negligible, manifestations of the companion almost undetectable in spectra.
		\item {\it Asymptotic- and red-giant branch star} -- significant eclipses will take place, colour would be shifted to red, the amplitude of light variations of the binary caused by pulsations would be significantly lower than in separate RR Lyrae, enrichment with heavier elements would be possible with AGB companion.
		\item {\it Horizontal branch star} -- significant eclipses will take place, possible colour excess should be detectable, lower amplitude of light changes than in separate RR~Lyrae will be observed.
	\end{itemize}
	
	In all these cases the confirmation of binarity via spectroscopy would be difficult because of low amplitudes in RV domain. Long-term tiny changes in RV can only be revealed using accurate template curves which are still missing \citep[see][this proceedings]{guggenberger2015}. Without them it is very difficult to find the zero points of systemic velocities from various measurements which are often based on different spectral lines. In the case of unavailable template curve there is also an indirect method through analysis of the scatter of the pulsation-phased RV curves. After removing the orbital motion the scatter should significantly decrease \citep[see fig. 7 in][]{liska2015a}.
\vspace*{-0.3cm}	
\section{Current situation in binary candidates}\label{Sec:Situation}
\begin{figure}[!ht]
\begin{center}	
\includegraphics[width=0.36\textwidth]{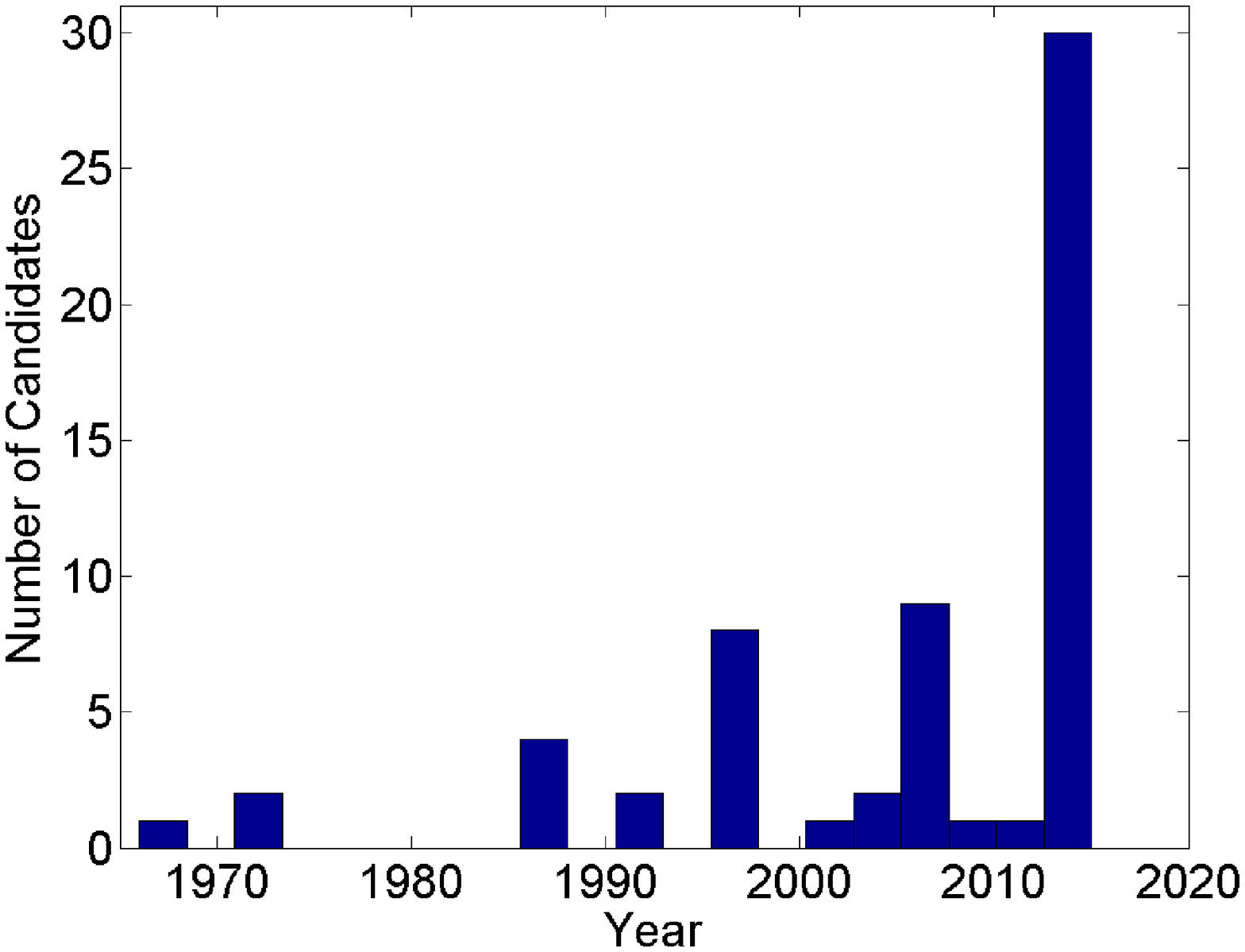}
\includegraphics[width=0.36\textwidth]{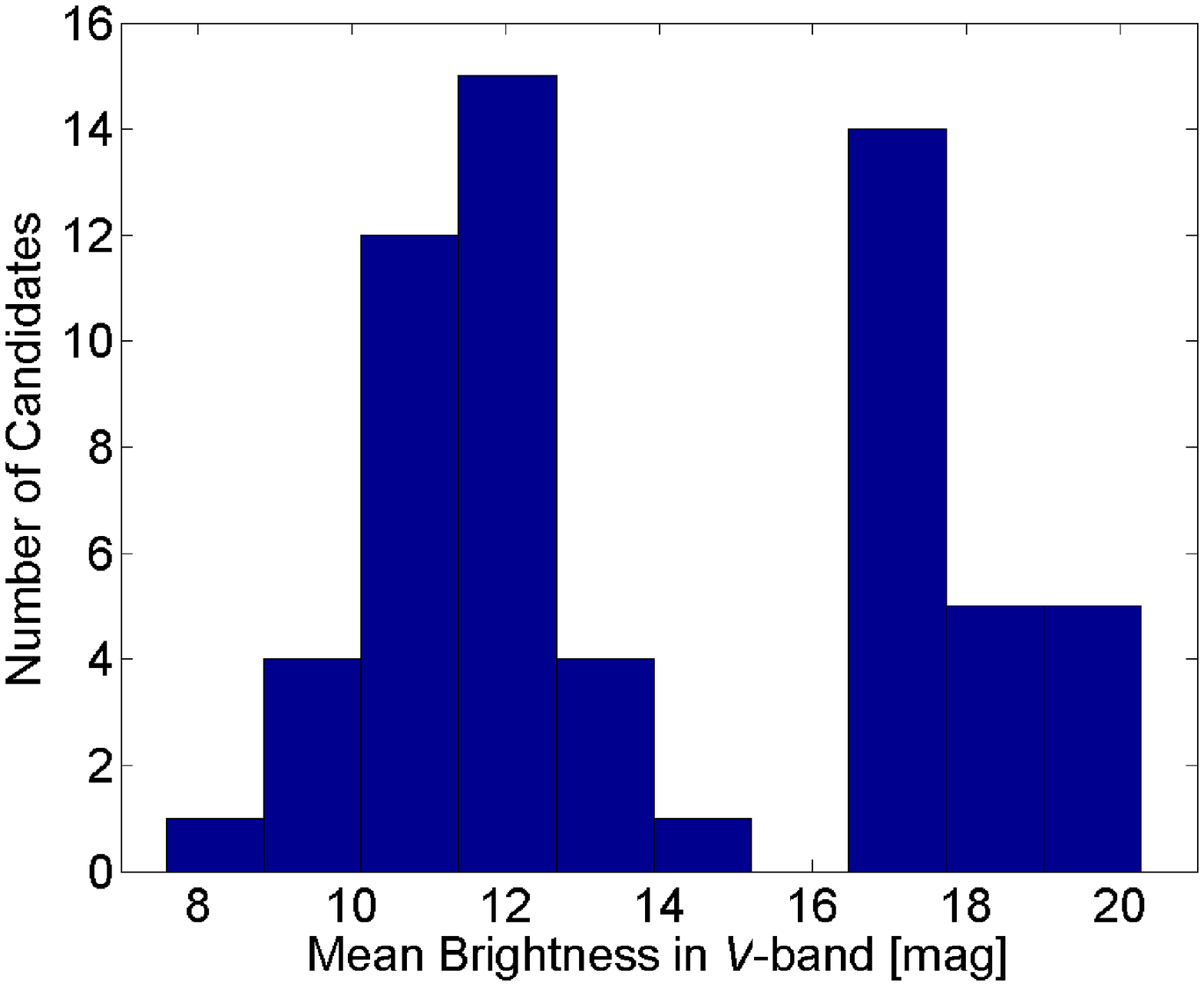}
\caption{Number of discovered candidates during last 50 years (the left panel) and magnitude distribution of known candidates.} 
\label{Fig:Year-Mag-hist} 
\end{center}
\end{figure}

\vspace*{-0.2cm}
	The first candidate for RR Lyrae in binary system was proposed in 1960s. However, the majority of candidates have been revealed only recently \citep{hajdu2015a,liska2015b}. The number of discovered candidates can be seen in the left panel of fig. \ref{Fig:Year-Mag-hist}. The mean-magnitude distribution of the candidates is bimodal due to selection effect (most of stars are either from the Galactic bulge, or bright stars from the Galactic field, the right panel of fig. \ref{Fig:Year-Mag-hist}). 
	
\begin{figure}[!ht]
\begin{center}	
\includegraphics[width=0.39\textwidth]{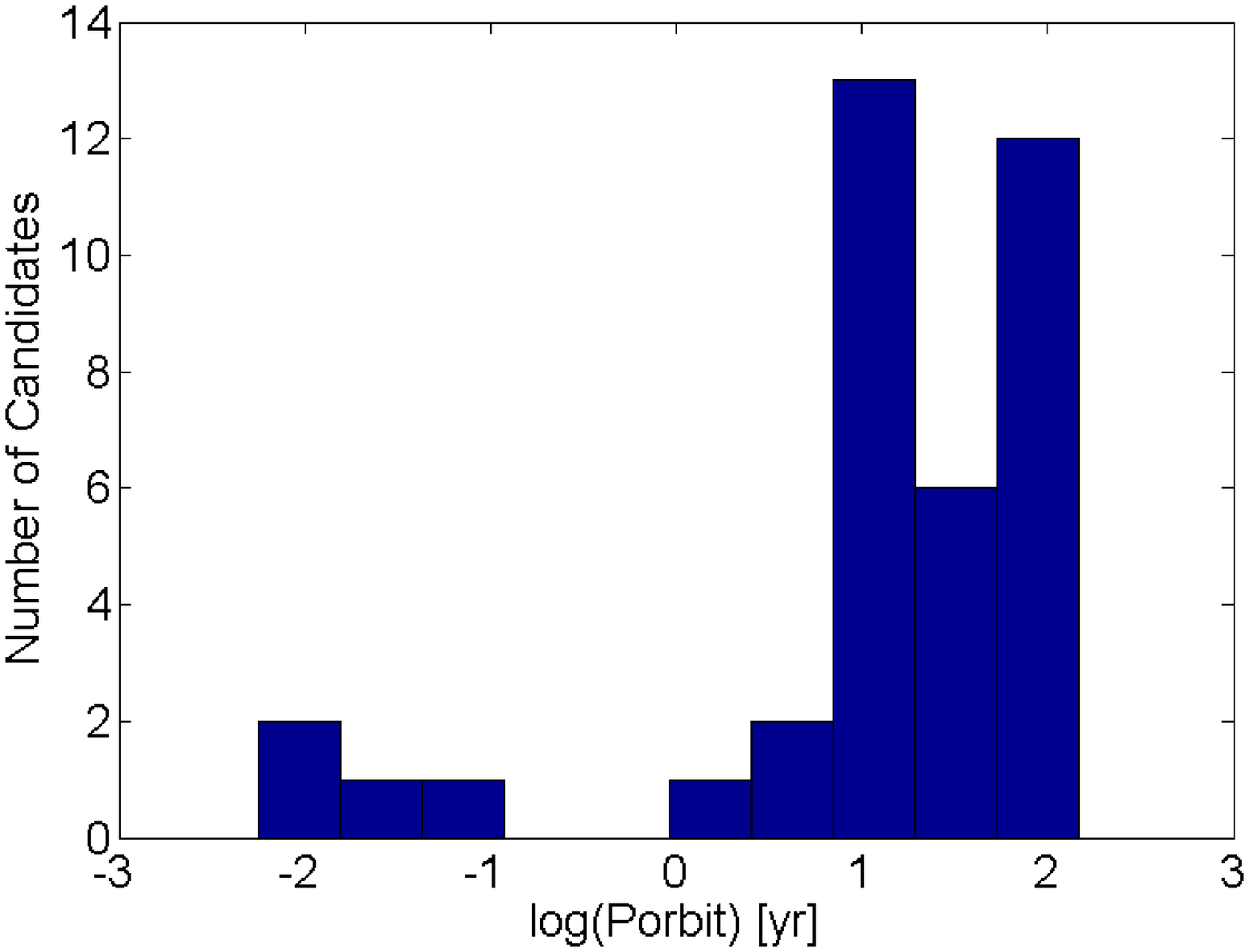}
\includegraphics[width=0.36\textwidth]{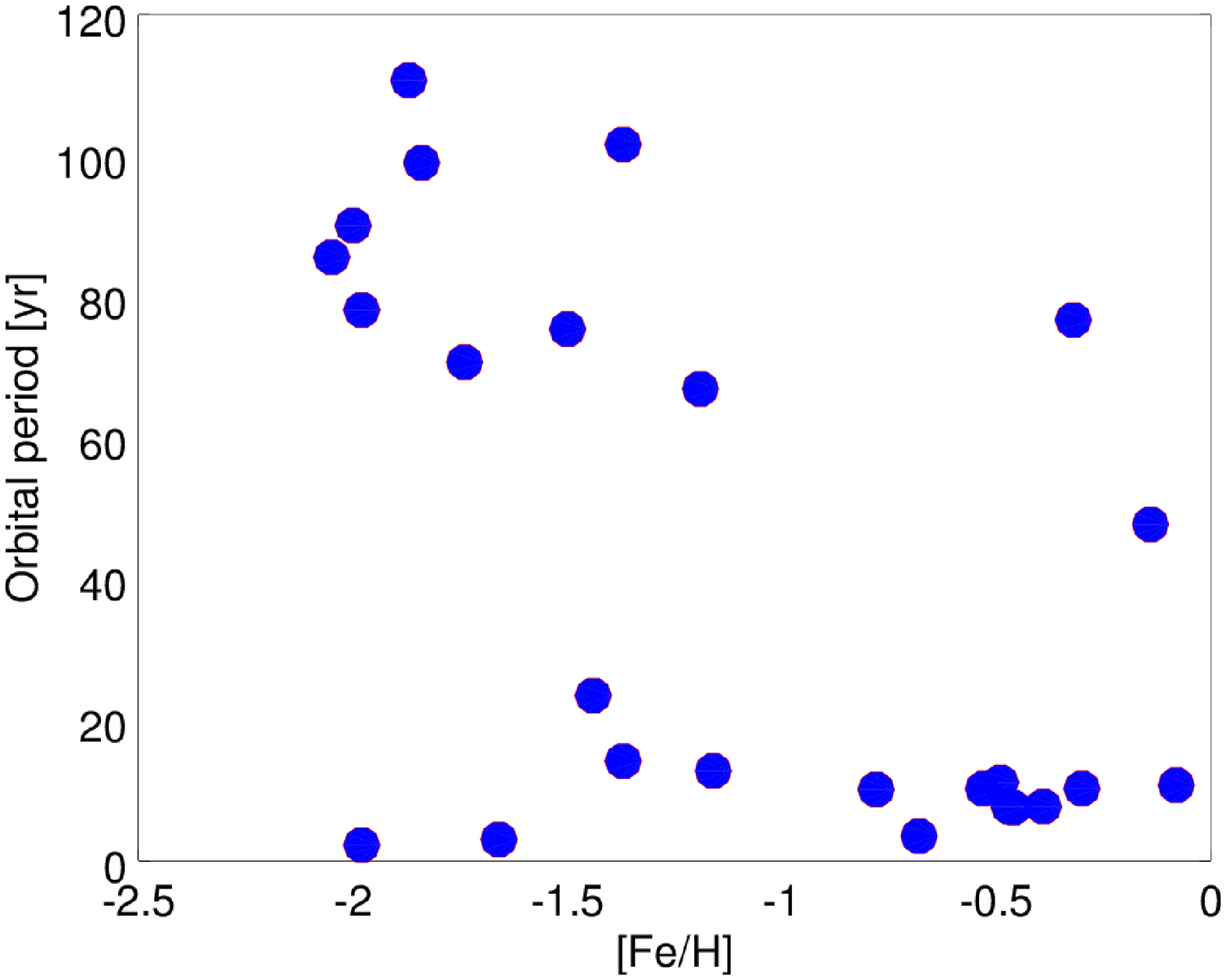}
\caption{Orbital period distribution (the left panel) and orbital period against metallicity (the right panel).} 
\label{Fig:Per-Metal-hist} 
\end{center}	
\end{figure}

	Except for a few candidates for short-period eclipsing binaries, all other stars have periods longer than a year (the left panel of fig. \ref{Fig:Per-Metal-hist}). Longer orbital periods mean that also semi-major axes are large, in the order of astronomical units or larger. When orbital periods are plotted against metallicity (the right panel of fig. \ref{Fig:Per-Metal-hist}), no apparent dependence is visible except for the splitting which is again an observation bias. About 1/5 of all candidates shows the Blazhko effect. Mass function in systems with models of the orbit ranges from $4\times10^{-6}$ to several tens of solar masses. For detail statistics see \citet[][this proceedings]{liska2015c}.
	
	Due to all discussed problems in sec. \ref{Sec:Characteristics} the most efficient method for revealing the candidates is looking for the variations caused by the orbital motion of RR Lyrae translating in cyclic period changes known as the Light-travel time effect (LiTE). Since this method is only indirect and the variations can be misclassified with other effects (secular erratic changes, long-term Blazhko effect etc.) an independent confirmation is needed in such objects.	

\vspace*{-1cm}

\vspace{-1.2cm}

\section{Interesting cases}\label{Sec:Cases}
\vspace*{-0.2cm}
	Among the galactic-field candidates, there are several very interesting cases which deserve attention. The first of them is TU UMa with orbital period of 23 years \citep{liska2015b} which is the shortest one among the long-period fraction of candidates. It is bright and therefore accurate template curve could be easily accessible. Confirmation of binarity should, therefore, be a task for a few next years.
	
	Period variations of two of the candidates indicate possible high mass companion -- RZ Cet ($M_{\rm 2,min}=1.15$\,M$_{\odot}$) and AT Ser ($M_{\rm 2,min}=1.9$\,M$_{\odot}$), two of the candidates, VX Her and RW Ari, showed suspicious decrease in brightness which might be caused by an unseen body \citep{fitch1966,wisniewski1971}. However, these events were unique and has never repeated. Assuming the depth of the decrease and colour measured in VX Her by \citet{fitch1966} the companion would be a horizontal-branch star. When we take the 83-yr orbital period \citep{liska2015b}, the eclipse should last 500\,d! 
	
	BB Vir and RS Boo were supposed to have extraordinary colour \citep{fitch1966,bookmeyer1977,kanyo1986} and possible LiTE was detected by \citet{liska2015b}.
\vspace*{-0.4cm}	
\section{Summary and future prospects}
\vspace*{-0.2cm}	 
	We discussed the status of the research in binary candidates with RR Lyrae component, their characteristics and problems with their detection. Parameters of possible component which would have crucial influence on observable characteristics were discussed. We also highlight the most interesting candidates.
	
	Only the future can prove whether the candidates are real binaries. It is proposed to focus on interesting objects with accumulated interesting features, for example eclipses and LiTE, in a systematic and long-term matter to get reliable results being capable to reveal tiny changes.
\vspace{-0.3cm}

\section*{Acknowledgements}
\vspace{-0.3cm}
Financial support of grants MUNI/A/1110/2014 and LH14300 is acknowledged. MS acknowledges the support of the postdoctoral fellowship programme of the Hungarian Academy of Sciences at the Konkoly Observatory as a host institution.
%\vspace*{-0.6cm}


\begin{thebibliography}{} 
\bibitem[Bookmeyer et al.(1977)]{bookmeyer1977} Bookmeyer, B.~B., 
Fitch, W.~S., Lee, T.~A., Wisniewski, W.~Z., \& Johnson, H.~L.\ 1977, RMxAA, 2, 235
\bibitem[Duquennoy\,\&\,Mayor(1991)]{duquennoy1991} Duquennoy, A., Mayor, M\ 1991 A\&A, 248, 485
\bibitem[Fitch et al.(1966)]{fitch1966} Fitch, W.~S., Wisniewski, W.~Z., \& Johnson, H.~L.\ 1966, Communications of the Lunar and Planetary Laboratory, 5, 3
\bibitem[Guggenberger et al.(2015)]{guggenberger2015}Guggenberger, E., Barnes, T. G., Kolenberg, K., 2015, this proceedings
\bibitem[Hajdu et al.(2015a)]{hajdu2015a} Hajdu, G., Catelan, M., Jurcsik, J., et al.\ 2015a, \mnras, 449, L113 
\bibitem[Hajdu (2015b)]{hajdu2015b} Hajdu, G., Catelan, M., Jurcsik, J. et al., 2015b, this proceedings
\bibitem[Jurcsik et al.(2012)]{jurcsik2012} Jurcsik, J., S\'{o}d\'{o}r, A., Hajdu, G. et al., 2012, MNRAS, 419, 2173
\bibitem[Kanyo(1986)]{kanyo1986} Kanyo, S.\ 1986, Commmunications of the Konkoly Observatory Hungary, 87, 1 
\bibitem[Kennedy et al.(2014)]{kennedy2014} Kennedy, C.~R., Stancliffe, R.~J., Kuehn, C., et al.\ 2014, \apj, 787, 6 
\bibitem[Lada(2006)]{lada2006} Lada, 2006, ApJ, 640, 63
\bibitem[Liakos et al.(2012)]{liakos2012} Liakos, A., Niarchos, P., Soydugan, E., Zasche, P., 2012, MNRAS, 422, 1250
\bibitem[Li\v{s}ka et al.(2015a)]{liska2015a} Liska, J., Skarka, M., Mikulasek, Z., Zejda, M., \& Chrastina, M.\ 2015a, arXiv:1502.03331 
\bibitem[Li\v{s}ka et al.(2015b)]{liska2015b} Liska, J., Skarka, M., Zejda, M., \& Mikulasek, Z.\ 2015b, arXiv:1504.05246
\bibitem[Li\v{s}ka\,\&\,Skarka(2015)]{liska2015c} Li\v{s}ka, J., Skarka, M., 2015, this proceedings (poster P-8)
%\bibitem[Mason et al. (2001)]{mason2001} Mason, B.\,D., Wycoff, G.\,L., Hartkopf, W.\,I. et al., 2001, AJ, 122, 3466
\bibitem[Pietrzynski et al.(2012)]{pietrzynski2012}Pietrzy\'{n}ski, G., Thompson, I.\,B., Gieren, W. et al. 2012, Nature, 484, 75
\bibitem[Sana\,\&\,Evans(2011)]{sana2011} Sana, H., Evans, Ch.\,J., 2011, IAUs, 272, 474
\bibitem[Szabados (2003)]{szabados2003} Szabados, L., 2003, IBVS, 5394, 1
\bibitem[Wi{\'s}niewski(1971)]{wisniewski1971} Wi\'{s}niewski, W.~Z.\ 1971, AcA, 21, 307 
\bibitem[Zhou(2014)]{zhou2014}Zhou, A.-Y. 2014, arXiv:1002.2729v5
\end{thebibliography}
\end{document}